\documentclass[reprint,amsmath,amssymb,aps]{revtex4-2}

\usepackage[modulo]{lineno}
\modulolinenumbers[10]
\usepackage{lipsum}
\usepackage{blindtext}
\usepackage{graphicx}
\usepackage{dcolumn}
\usepackage{bm}
\usepackage[hidelinks]{hyperref}
\usepackage{xcolor}

\usepackage{ulem}
\usepackage[dvipsnames]{xcolor} 
\usepackage{multirow}
\usepackage{array}
\usepackage{siunitx}
\usepackage{braket}

\usepackage{soul}

\begin{document}

\preprint{APS/123-QED}

\title{Dual role of core electrons in electronic friction}

\author{Runfeng Zhou$^{1}$}\email{rz391@cam.ac.uk}
\author{Emilio Artacho$^{1,2,3}$}\email{ea245@cam.ac.uk}

\affiliation{
$^1$Theory of Condensed Matter, Cavendish Laboratory, 
University of Cambridge, J. J. Thomson Avenue, Cambridge CB3 0US, United Kingdom\\
$^2$CIC Nanogune BRTA and DIPC, Tolosa Hiribidea 76, 20018 San Sebastian, Spain\\
$^3$Ikerbasque, Basque Foundation for Science, 48011 Bilbao, Spain}

\begin{abstract}
  Non-equilibrium energy dissipation in multi-shell swift-ion/matter systems 
remains a fundamental yet incompletely understood problem, with electronic 
stopping power \(\mathcal{S}_\text{e}\) as a relevant observable for electronic 
friction.
  Using real-time time-dependent density functional theory, we perform 
first-principles calculations of \(\mathcal{S}_\text{e}\) for beryllium 
self-irradiation with explicit treatment of all electrons. 
  Our results reveal a Bragg peak exhibiting a distinct structure which 
lies beyond the reach of standard models.  
  We attribute its appearance to a dual effect of the presence of core 
electrons, by which their excitation provides an additional dissipation 
channel while simultaneously suppressing valence electron excitations,
contrary to earlier proposals.
  Electron capture by the projectile's core from the host cores 
is behind such suppression rather than Pauli blocking. 
\end{abstract}

\maketitle

  Non-equilibrium dissipative processes of ions shooting through matter 
represent a central topic in physics and carry significant 
technological implications, including applications in medical ion 
therapy~\cite{maradia2025accelerator,deng2005beyond,shepard2023electronic}, 
advanced energy conversion systems~\cite{roy2024engineered}, space radiation
shielding~\cite{kodaira2024space}, and microelectronic chip 
fabrication~\cite{moll2018focused}. 
  As ions traverse a material, they collide with the atomic nuclei and 
electrons of the host, losing kinetic energy and transferring it to the medium. 
  When the ion velocity approaches characteristic electron velocities 
(\(\gtrsim\) 1 atomic unit, a.u.), energy dissipation is dominated by 
interactions between the projectile and the host electrons~\cite{correa2018calculating}. 
  The energy loss per unit path length is quantified as the electronic stopping 
power, \(\mathcal{S}_\text{e} = -\text{d}E/\text{d}x\)~\cite{massillon2025anomalies}. 
  It initially increases with ion kinetic energy or velocity (\(\mathcal{S}_\text{e}/v\) 
defines the friction coefficient), reaches a maximum known as the Bragg 
peak~\cite{brown2004centenary,matias2025stopping}, and subsequently decreases, 
as electrons become too slow to respond.
  There is a long history of $S_e$ calculations based on various
techniques~\cite{Sigmund2006particle}, and it has been calculated from first 
principles reasonably accurately for various systems~\cite{correa2018calculating}.
  However, as the electronic shell structures of both the ions and host materials 
become increasingly complex, the behavior of the electronic stopping power 
becomes harder to understand and simulate~\cite{ullah2018core}.

  \textit{In silico} approaches based on time-dependent density functional theory 
\cite{runge1984density} (TDDFT), in particular real-time TDDFT (rt-TDDFT), have 
the capability of directly analyzing microscopic ultrafast non-equilibrium electron 
dynamics~\cite{marques2004time,ullrich2025snapshot}. 
  In contrast to experimental approaches such as the energy-loss method, which 
typically yield only averaged energy transfer over a finite path 
length~\cite{primetzhofer2011electronic}, rt-TDDFT enables a direct, time-resolved 
evaluation of electronic stopping power and related magnitudes such as particle 
density, and single particle Kohn-Sham (KS) functions.
  As a result, rt-TDDFT has been successfully applied across a broad range of 
projectile energies, from low-energy ions such as H and He~\cite{zeb2012electronic} 
to high-energy ions such as Ni~\cite{ullah2018core}, and to diverse target 
materials, including crystalline metals~\cite{zeb2012electronic,sun2025electron}, 
insulators~\cite{pruneda2007electronic,qi2022ab,massillon2025anomalies,gao2025unveiling} 
and semiconductors~\cite{lim2016electron}, and also amorphous water~\cite{yao2019k,
shepard2023electronic,matias2025stopping} and warm dense 
matter~\cite{kononov2024reproducibility}.

  Unlike at low projectile velocities, core electrons of both the host and the 
projectile play a significant role in non-adiabatic dissipation during 
swift ion motion. 
  A series of studies have extended the scope beyond valence electrons to 
include semicore and even core electrons, covering light ions in all-electron 
hosts~\cite{yost2017examining,yao2019k}, projectiles with core 
electrons~\cite{ojanpera2014electronic}, and systems in which both the 
projectile and host contain semicore electrons~\cite{ullah2018core}, 
revealing that the explicit loss of projectile core electrons exposes a 
stronger ionic potential~\cite{ojanpera2014electronic,ullah2018core}, while 
host core electrons provide additional channels for energy 
transfer~\cite{yost2017examining,yao2019k,ullah2018core}, all of which 
enhance stopping. 
  Moreover, the channels are not independent: Yao \textit{et al.} reported 
that core electrons can further enhance valence excitation in water via the 
shake-up effect~\cite{yao2019k}.

  Beryllium (Be) metal is of interest both in equilibrium~\cite{bista2026real} and out of it as target of irradiation. It  
combines extremely low density, exceptional stiffness, and excellent 
thermal stability, while its low atomic number provides favorable neutronic 
properties~\cite{foley2017beryllium} important for aerospace, optical, 
electronic, and nuclear technologies. 
  In these energetic environments, primary knock-on atoms initiate 
self-irradiation, a fundamental mechanism governing radiation-induced 
degradation~\cite{jackson2016simulations}. 
  However, a knowledge gap persists due to sparse  experimental stopping 
power data for the Be-in-Be system.
  Its electronic structure with two core electrons and two 
valence electrons per atom makes it an ideal system to further our 
understanding of multi-shell electronic friction.

  In this paper, we employ rt-TDDFT to investigate the prototypical metallic 
self-irradiation system, a Be projectile in bulk Be. 
  The study enables a fully explicit treatment of all electronic shells of 
both projectile and host, including the 1$s$ core electrons. 
  By selectively sampling characteristic trajectories, we naturally decouple 
valence-dominated excitations from intense core-core interactions within a 
physically consistent framework without resorting to artificial pseudopotential 
modifications, thereby closing a long-standing gap in the microscopic 
understanding of electronic friction.


\begin{figure}[t]
    \centering
    \begin{minipage}{0.49\textwidth}
    \centering
    \includegraphics[width=\linewidth]{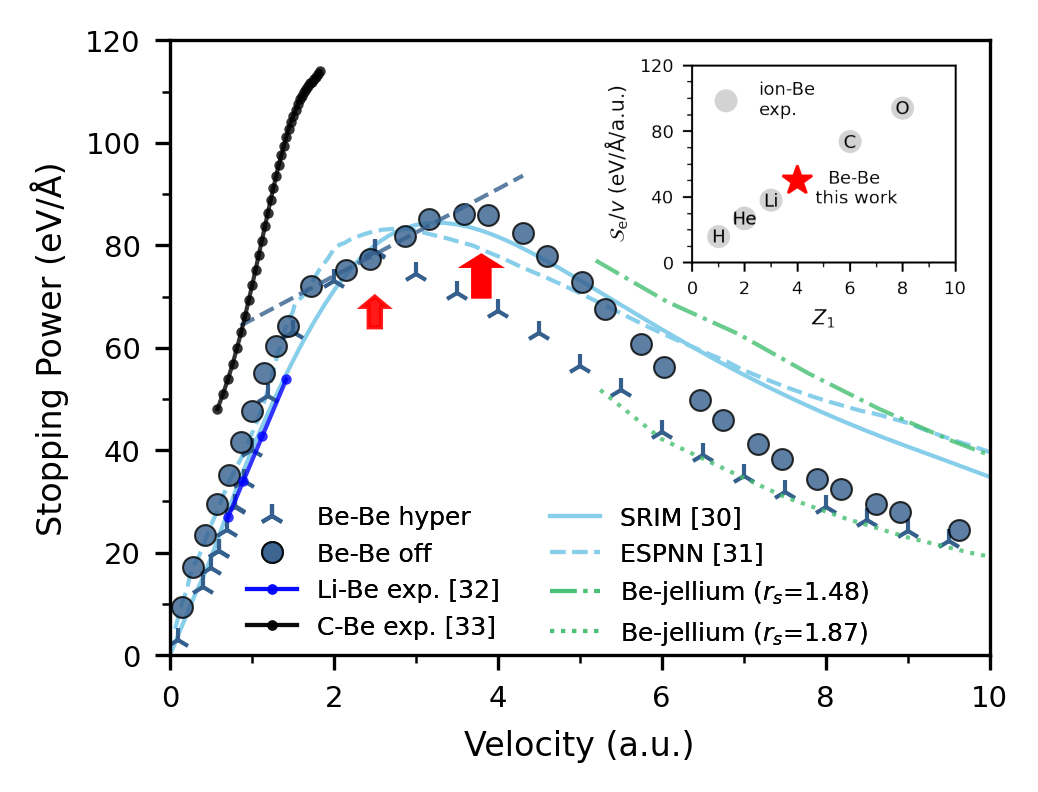}
    \end{minipage}
    \caption{\label{fig1} Electronic stopping power \(\mathcal{S}_\text{e}\) 
    for a Be projectile in bulk Be host as a function of velocity. 
    Filled circles and inverted Y-shapes represent \(\mathcal{S}_\text{e}\) 
    along the off-channeling and hyper-channeling trajectories, respectively. 
    Skyblue solid and dashed lines show predictions of \(\mathcal{S}_\text{e}\) from the 
    SRIM~\cite{ziegler2010srim} and ESPNN~\cite{bivort2022espnn} models. 
    Green dotted and dash-dotted lines show \(\mathcal{S}_\text{e}\) for a Be 
    projectile in dilute and dense jellium. 
    Blue-green dashed line serves as a guide to the eye indicating the slope 
    change near the Bragg peak in off-channelling. 
    Blue and black dots represent experimental stopping power data for Li in Be
    \cite{neuwirth1975electronic} and C in Be \cite{santry1991measured}, respectively.
    The red arrows indicate the positions of Bragg peaks in the hyper- and 
    off-channeling trajectories.
    Inset: Gray dots show experimental friction coefficient values for 
    different projectiles in Be in the low-velocity regime plotted against 
    atomic number, $Z_1$, where each point represents an independent 
    experimental dataset~\cite{brunner1980stopping,santry1980stopping,
wrean1994total,neuwirth1975electronic,santry1991measured,santry1992measured,
    chu1968range}; the red pentagram shows the simulation result in this work.
    }
\end{figure}



  The rt-TDDFT calculations were performed using the plane-wave pseudopotential 
based INQ/QB@LL program~\cite{andrade2021inq,draeger2017massively}.
  The Be host was modeled as a \(5 \times 6 \times 3\) hexagonal close-packed 
supercell containing 180 Be atoms in periodic boundary conditions, on 
a unit cell defined by $a=b=$ 2.285 \AA, $c=$ 3.585 \AA, and 
$\alpha=\beta=90^{\circ}$ and $\gamma=120^{\circ}$.
  A Be projectile was introduced at a threefold hollow site and propagated 
through the host at constant velocity \(v\), while the host atoms were frozen 
to isolate the pure electronic stopping process. 
  All the electrons, core and valence, were considered explicitly, and a 
norm-conserving Vanderbilt pseudopotential~\cite{hamann2013optimized} 
was used to remove the potential singularity at nuclear positions.
  Exchange-correlation effects were treated using the Perdew-Burke-Ernzerhof 
generalized gradient approximation~\cite{perdew1996generalized}. 
  The KS wavefunctions were expanded in a basis of plane waves of kinetic 
energy up to 50 Ry and explicitly propagated in time, using \(\Gamma\)-point sampling. 
  \(\mathcal{S}_\text{e}\) was determined from a linear fit of total energy 
versus projectile displacement.

  Although direct experimental data for Be ions in Be are unavailable, 
existing low-velocity experimental \(\mathcal{S}_\text{e}\) data for other 
ion projectiles in Be \cite{brunner1980stopping,santry1980stopping,wrean1994total,
neuwirth1975electronic,santry1991measured,santry1992measured,chu1968range} 
provide useful reference for validation. 
  Our calculation results are reasonably located between the reported values 
for Li in Be~ \cite{neuwirth1975electronic} and C in Be~ \cite{santry1991measured}. 
  The low-velocity experimental values of \(\mathcal{S}_\text{e}/v\), 
i.e., the friction coefficient, are plotted against the atomic number \(Z_1\) 
in the inset of Fig.~\ref{fig1}, with our friction coefficient for Be (marked
as a red pentagram) lying well within what expected from interpolation following 
the observed linear behavior for the shown $Z_1$ range \footnote{The $Z_1^2$ 
dependence expected from linear response is known to be modified by non-linear 
effects giving rise to, e.g., the well-known $Z_1$ oscillations with a first 
maximum normally at higher $Z_1$ values than the ones presented.}.

  To elucidate the impact of core electrons on energy dissipation, two 
representative trajectories were considered: a high-symmetry hyper-channeling 
path along the \(c\) axis and a random-equivalent off-channeling 
trajectory~\cite{kononov2023trajectory}.
  These choices represent two physical limits: the former minimizes core-core 
overlap thereby isolating valence-dominated interactions, whereas the latter 
provides a realistic account of the strong interactions encountered in close 
collision events affecting the core regions. 
  The contrast between these trajectories enables a direct quantification of 
core-level contributions to \(\mathcal{S}_\text{e}\).

  The calculated \(\mathcal{S}_\text{e}\) in both off- and hyper-channeling 
qualitatively follow the characteristic non-monotonic dependence on projectile 
velocity \(v\): increasing linearly at first, reaching a maximum (the Bragg peak), 
and subsequently decreasing (see Fig.~\ref{fig1}).
  But they exhibit pronounced differences, particularly in the vicinity of 
the Bragg peak. 
  In the low-velocity (linear) regime, \(\mathcal{S}_\text{e}\) shows nearly 
identical values for both trajectories, with a slight enhancement in off-channeling.
  However, the structure (position, magnitude, and even the shape) of the 
respective Bragg peaks display pronounced differences (indicated by red arrows 
in Fig.~\ref{fig1}).
  For hyper-channeling it is relatively simple, appearing at \(v \approx\) 
2.5 a.u. with \(\mathcal{S}_\text{e} \approx \) 79 eV/Å. 
  In contrast, the Bragg peak in off-channeling displays a more complex 
two-stage mode: \(\mathcal{S}_\text{e}\) initially reaches a turning point 
at \(v \approx\)~2.0 a.u., after which it continues to increase at a reduced rate, 
exceeds the hyper-channeling peak, and eventually reaches the true Bragg peak 
at \(v \approx 3.8\)~a.u. with \(\mathcal{S}_\text{e} \approx \) 86 eV/Å 
(see the blue-green dashed line in Fig.~\ref{fig1}).
  Beyond the Bragg peak, this discrepancy of \(\mathcal{S}_\text{e}\) 
persists and only gradually diminishes as the velocity approaches 
$\sim  10$ a.u.
  The distinctive Bragg peak structure indicates a significant contribution 
of core electrons to \(\mathcal{S}_\text{e}\) in the random trajectory, pointing 
to a richer multi-channel mechanism for energy transfer and dissipation.

  To place the special Bragg peak features in context, the calculated 
\(\mathcal{S}_\text{e}\) is compared in Fig.~\ref{fig1} with the semi-empirical 
SRIM~\cite{ziegler2010srim} and machine-learned ESPNN~\cite{bivort2022espnn} models. 
  The calculation results show good agreement with the theoretical predictions 
in the low-velocity regime. 
  However, the complex Bragg peak structure observed above is not reproduced 
by either model. 
  Both models describe only a simple peak structure. 
  Specifically, SRIM predicts a peak at \(v \approx\) 3.3 a.u. with 
\(\mathcal{S}_\text{e} \approx \) 84.4 eV/Å and ESPNN predicts a peak at 
\(v \approx\) 3.0 a.u. with \(\mathcal{S}_\text{e} \approx \) 83.3 eV/Å. 
  Although the peak magnitudes are close to the calculation, their positions 
and shapes deviate significantly. 
  At velocities beyond 5 a.u. both models deviate from our first-principles results.
  This discrepancy is discussed in the Appendix.

\begin{figure}[htbp]
    \centering
    \includegraphics[width=\linewidth]{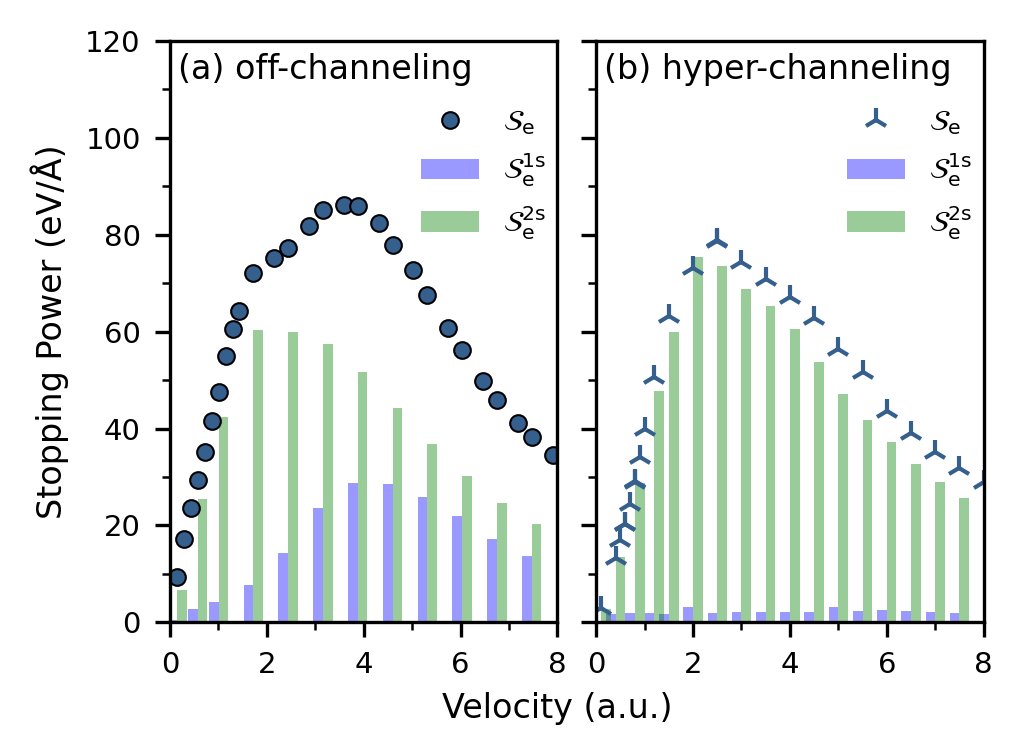}
    \caption{\label{fig2} Electronic stopping power \(\mathcal{S}_\text{e}\) 
    (circles and inverted Y-shape) and its decomposed components, 1$s$ 
    (\(\mathcal{S}_\text{e}^{1s}\), blue bars) and 2$s$ (\(\mathcal{S}_\text{e}^{2s}\), 
    green bars), versus projectile velocity along (a) off-channeling and 
    (b) hyper-channeling trajectories. 
    } 
\end{figure}

  To understand the mechanism underlying the formation of the distinctive 
Bragg peak's shape, we study the contributions to stopping from core and 
valence electrons defined as \(\mathcal{S}_\text{e}^{1s} = \text{d}
\varepsilon_{1s}/\text{d} x \) and \(\mathcal{S}_\text{e}^{2s} = 
\text{d}\varepsilon_{2s}/\text{d} x \) (see Fig.~\ref{fig2}), where 
\( \varepsilon_{1s} = \sum_{i \in {1s}} f_i \bra{\psi_i(t)} \hat{h}_\text{KS} 
\ket{\psi_i(t)} \) and \( \varepsilon_{2s} = \sum_{i \in {2s}} f_i 
\bra{\psi_i(t)} \hat{h}_\text{KS} \ket{\psi_i(t)}\) denote the sums of 
the energy expectation values of KS wavefunctions associated with the 
1$s$ and 2$s$ bands (see Fig.~\ref{fig-a2}), and \( f_i \) the corresponding 
occupation numbers. 
  The sum of \(\mathcal{S}_\text{e}^{1s}\) and \(\mathcal{S}_\text{e}^{2s}\) 
reproduces the total \(\mathcal{S}_\text{e}\) within 8\% (with the remaining 
discrepancy due to double-counting terms), thereby providing a reliable analytical 
tool.


  The decomposed stopping components as a function of \(v\) presented in 
Fig.~\ref{fig2} reveal a stark contrast between hyper-channeling and 
off-channeling trajectories.
  The contribution of core electrons is barely noticeable in hyper-channeling 
but plays a significant role in off-channeling, simultaneously enhancing 
and shifting the Bragg peak. 
  Specifically, the core component in off-channeling reaches a peak at 
\(v \approx 4\) a.u. with \(\mathcal{S}_\text{e}^{1s} \approx 29~\text{eV/Å}\), 
while the valence component peaks at \(v \approx 2\) a.u. with
\(\mathcal{S}_\text{e}^{2s} \approx 60~\text{eV/Å}\). 
  The resulting velocity mismatch between the individual 1$s$ and 2$s$ 
Bragg peak maxima gives rise to the distinct two-stage increased and shifted 
Bragg peak structure observed in off-channeling. 
  Near the Bragg peak, \(\mathcal{S}_\text{e}^{1s}\) accounts for up to 36\% 
of the total \(\mathcal{S}_\text{e}\), reaching a maximum of 45\% at higher 
velocities. 
  This substantial core dissipation channel, cooperating with the valence 
channel, significantly enhances the overall \(\mathcal{S}_\text{e}\).  

  Particularly noteworthy is that the opening of the core channel 
simultaneously exerts a suppressive effect on the valence channel. 
  This is manifested in the fact that \(\mathcal{S}_\text{e}^{2s}\) 
in off-channeling falls markedly below that in hyper-channeling at 
velocities above 1 a.u., with a reduction exceeding 20\% around the peak 
region.
  It reveals a previously unknown dual effect of core electrons, both 
cooperating and countering the stopping: while they enhance the total 
stopping power, they concurrently suppress the valence contribution. 
  This behavior stands in clear contrast to Ref. \cite{yao2019k}, where 
core electrons were reported to promote valence excitation via a shake-up 
effect in the case of proton in liquid water. 


\begin{figure}[htbp]
    \centering
    \begin{minipage}{0.49\textwidth}
    \centering
    \includegraphics[width=\linewidth]{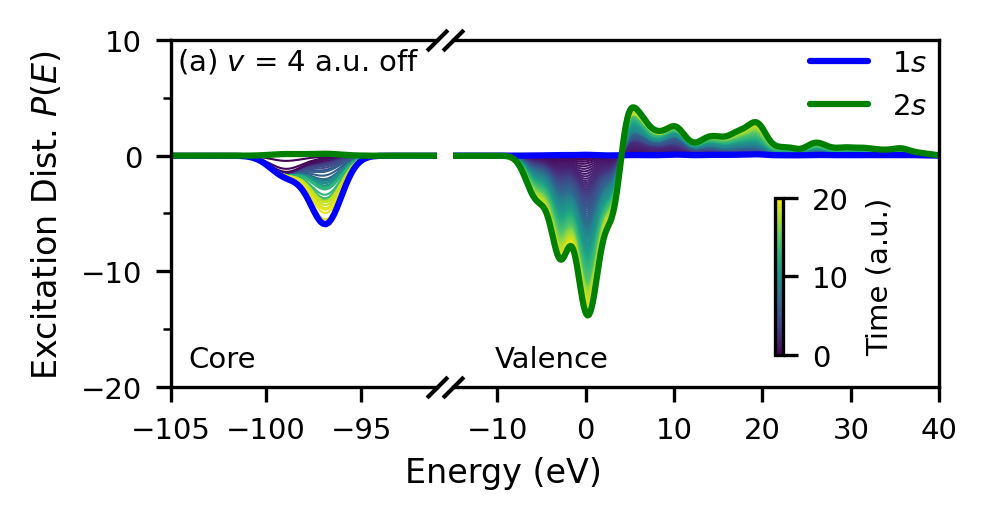}
    \end{minipage}\hfill
    \begin{minipage}{0.49\textwidth}
    \centering
    \includegraphics[width=\linewidth]{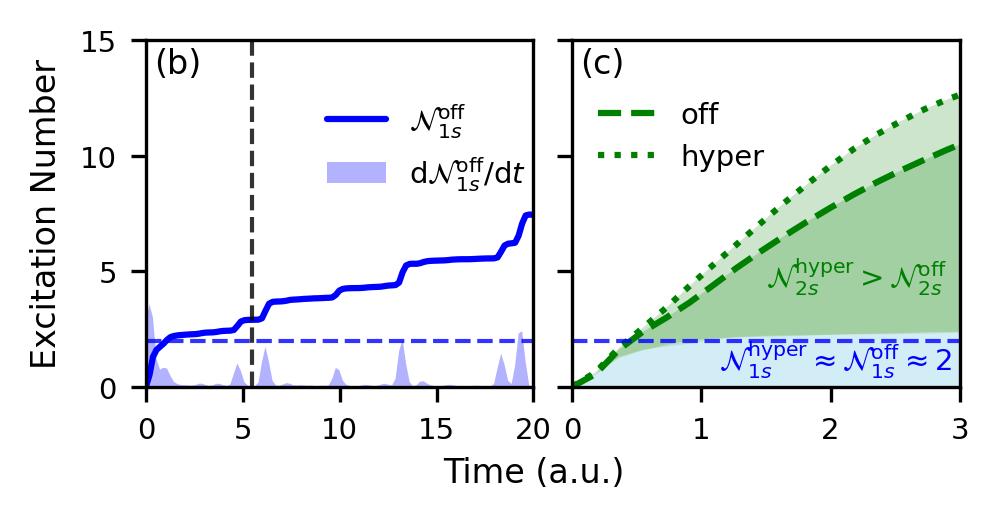}
    \end{minipage}
    \caption{\label{fig3} 
    Time evolution of the excitation distribution. 
    (a) Excitation distribution \(P(E)\) at \(v =\) 4 a.u. in off-channeling. 
    Color gradation: lighter colors correspond to larger time; 
    Gaussian broadening \(\sigma = \) 0.8 eV. 
    Blue and green solid lines show \(P_{1s}(E)\) and \(P_{2s}(E)\) at \(t = 20\)  
    a.u., respectively.
    (b) Excitation number of the 1$s$ bands, \(\mathcal{N}_{1s}\) (integral 
    of $P_{1s}(E)$ over relevant range, see Appendix). 
    In skyblue \(\text{d}\mathcal{N}_{1s}/\text{d}t\).
    Its peaks indicate core-electron excitation events. 
    Blue dashed line denotes projectile core ionization. 
    Black dashed line marks the completion of the first pronounced host 
    core excitation. 
    (c) Early-stage excitation number of the 1$s$ (blue) and 2$s$ (green) 
    bands in hyper- and off-channeling (dashed and dotted lines, respectively).
    } 
\end{figure}


  A possible intuitive explanation of the duality is that core 
electron excitations partially occupy high-energy states otherwise available 
to the 2$s$ electrons. 
  Due to the Pauli exclusion principle, this effectively blocks 2$s$ transitions. 
  Figure~\ref{fig3}(a) presents the excitation distribution for off-channeling at 
\(v =\) 4.0 a.u.
  The number of excited core electrons \(\mathcal{N}_{1s} \), estimated from 
the depletion of the 1$s$ band [\( P(E,t) < 0 \)] (see Appendix for definitions 
and details), accounts for \(\sim\)15\% of the total excited electrons 
(\(\mathcal{N}_{1s}+\mathcal{N}_{2s}\)), consistent with the \(\sim\)20\% 
reduction in \(\mathcal{S}_\text{e}^{2s}\). 

  Although this blocking hypothesis appears reasonable, it is insufficient. 
  First, by decomposing the excitation distributions of the 1$s$ and 2$s$ bands 
(blue and green solid lines in Fig.~\ref{fig3}(a)), we observe that the excitation 
of core electrons to states slightly above the Fermi-level is extremely low, 
typically one to two orders of magnitude smaller than that of valence excitations. 
  Second, by tracking \(\mathcal{N}_{1s}\) (blue solid line in Fig.~\ref{fig3}(b)), 
we find that 1$s$ excitations occur as discrete events (skyblue peaks, 
\(\text{d}\mathcal{N}_{1s}/\text{d}t\)), requiring extremely close nuclear 
encounters that promote electrons to high-energy states up to approximately 
140 eV on average. 
  This value is estimated from the ratio \(\Delta\varepsilon_{1s}/\Delta\mathcal{N}_{1s}\).  
  Such an energy scale lies far above that of the 2$s$ excitations near Fermi level.
  These results indicate that the two channels are well separated in energy 
and therefore do not compete via Pauli exclusion.
  Note that a small fraction of 2$s$ electrons relax into core level (in an 
Auger-like process), seen as a very small bump in the green curve 
(Fig.~\ref{fig3}(a)), which affects \(\mathcal{S}_\text{e}^{2s}\) by at most 
0.5 eV/Å and is therefore negligible.

  Moreover, as shown in Fig.~\ref{fig3}(b), the initial stage in 
the evolution of \(\mathcal{N}_{1s}\) shows that only the two core electrons 
of the projectile are excited, followed by a period during which no 
further 1$s$ excitation occurs. 
  This feature enables us to isolate and analyze the decoupled contribution 
of 1$s$ excitation in the early stage. 
  As further demonstrated in Fig.~\ref{fig3}(c), the suppressive effect of core 
electrons can arise (indicated by a substantially smaller \(\mathcal{N}_{2s}\) 
in off-channeling) even without host core excitation (nearly identical 1$s$ 
transitions, blue region with \(\mathcal{N}_{1s} \approx 2\)).


\begin{figure}[htbp]
    \centering
    \begin{minipage}{0.5\textwidth}
    \centering
    \includegraphics[width=\linewidth]{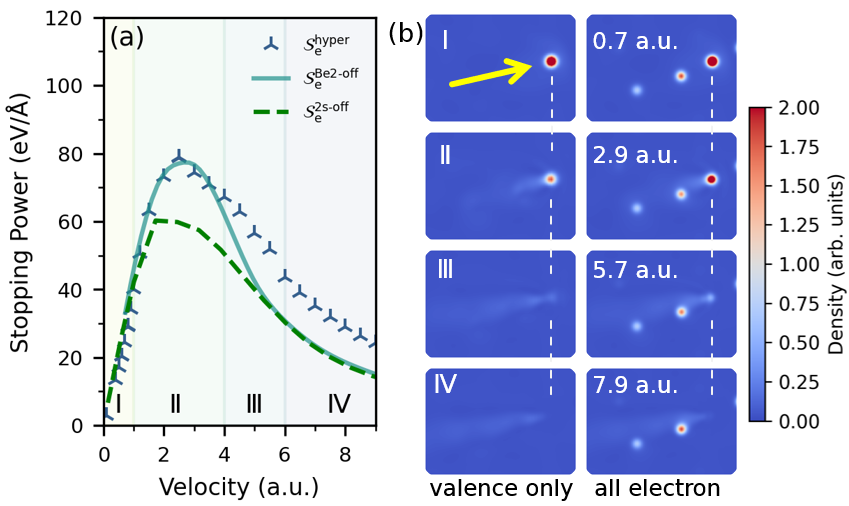}
    \end{minipage}
    \caption{\label{fig4} Four velocity regimes and the screening effect in 
    quasi-valence-electron systems. 
    (a) Stopping power of three quasi-valence-electron systems: the all-electron 
    hyper-channeling case \(\mathcal{S}_\text{e}^\text{hyper}\), 2$s$ component 
    of the all-electron off-channeling case \(\mathcal{S}_\text{e}^{2s\text{-off}}\), 
    and the pseudized off-channeling case \(\mathcal{S}_\text{e}^\text{Be2-off}\) 
    (pseudized host Be atoms and all-electron Be projectile). 
    (b) Electron capture from host cores, indicated by the difference in electron 
    density in off-channeling with all-electron and valence-only (pseudized) host system. 
    The yellow arrow points the trajectory, and the white dashed lines mark the ion 
    position after a displacement of 21.89 Bohr, corresponding to the same position 
    immediately following a core-collision event shown by the black dashed line 
    in Fig.~\ref{fig3}(b).  } 
\end{figure}

  Since the suppressive effect is independent of the host core electron 
excitation process, we attribute it to the charge screening induced by 
core electrons. 
  An experimentally representative trajectory traverses a higher proportion 
of low-valence-density core-electron regions, where valence electrons are 
sparse, thereby reducing the valence contribution. 
  The characteristic radius of core-electron region is \(\sim\)1 a.u., 
occupying more than 10\% of the system volume, consistent with the observed 
reduction in \(\mathcal{S}_\text{e}^{2s}\). We refer to this as 
\textit{static screening}.

  To further clarify the screening mechanism, we investigate an auxiliary 
system in which an all-electron Be projectile traverses a Be host with 
frozen (pseudized) core electrons (denoted as Be2) along the 
off-channeling trajectory. 
  This pseudopotential preserves the screening of the nucleus by host 
core electrons without allowing host core excitations.
  Figure~\ref{fig4}(a) compares the stopping power of three quasi-valence-electron 
systems: the all-electron hyper-channeling case \(\mathcal{S}_\text{e}^\text{hyper}\), 
the 2$s$ component of the all-electron off-channeling case 
\(\mathcal{S}_\text{e}^{2s\text{-off}}\), and the pseudized off-channeling 
case \(\mathcal{S}_\text{e}^\text{Be2-off}\). 
  In the low-velocity regime, all three results coincide; near the peak region, 
\(\mathcal{S}_\text{e}^\text{Be2-off}\) and \(\mathcal{S}_\text{e}^\text{hyper}\) 
are comparable and both exceed \(\mathcal{S}_\text{e}^{2s\text{-off}}\); 
as the velocity increases, \(\mathcal{S}_\text{e}^\text{Be2-off}\) decreases 
and converges toward \(\mathcal{S}_\text{e}^{2s\text{-off}}\), with both 
remaining below \(\mathcal{S}_\text{e}^\text{hyper}\).

  Figure~\ref{fig4}(b) shows the velocity-dependent evolution of the 
electron density as the ion traverses a core region. 
  With increasing velocity, the electronic cloud transitions from a
quasi-adiabatic, nearly spherical distribution to a distorted configuration 
characterized by a reduced frontal density and the emergence of a comet-like 
tail, eventually losing the frontal accumulation entirely. 
  Notably, the figure shows that the projectile has a higher electron 
density in the core region for the all-electron host system, indicating 
that the projectile captures extra electrons from the host cores.   
  We attribute it primarily to the alignment of projectile/host core-level 
energies.
  Thus, the lower effective charge leads to stronger \textit{dynamic screening} 
and lower valence stopping power.
  
  These results allow us to classify the role of core electrons into three 
velocity-dependent mechanisms, manifested across four regimes: a quasi-adiabatic 
regime (I), a Bragg regime (II), a crossover regime (III), and a high-velocity 
perturbative regime (IV).
  In regime I, host core electrons remain essentially inactive, while valence 
electrons are excited and produce a linear (Stokes-like) dissipation of the 
projectile.
  In regime II, electron capture leads to enhanced dynamic screening of the 
projectile, thereby reducing the effective valence contribution to stopping.
  In regime III, electron capture decreases due to insufficient interaction 
time, making a gradual transition from dynamic to static screening.
  In regime IV, the swift-ion/matter interaction becomes perturbative, 
where electron capture is negligible and static screening dominates, 
and thus suppresses the valence contribution.

  In summary, electronic stopping power calculations that explicitly 
include all electrons have been performed for Be self-irradiation. 
  The key Bragg peak features are analyzed in detail. 
  Explicit projectile core electrons allow for enhanced electron 
delocalization, increasing the effective ionic potential and thus the 
stopping power while shifting the Bragg peak. 
  Explicit host core electrons permit the excitation of 1$s$ electrons, 
providing an additional dissipation channel. 
  Energy decomposition analysis demonstrates that core electrons 
exhibit a cooperative-counteractive duality, where they enhance the 
overall stopping while simultaneously suppressing excitation in other 
channels.
  The analysis shows they do so by dynamic and static screening the 
valence electron response instead of occupying high-energy states, 
leading to the observed multi-stage evolution of the Bragg peak. 
  This work elucidates the multi-channel coupling mechanism in 
electronic stopping and provides a foundational framework for 
analyzing the energy dissipation in more complex systems.
  

  \textit{Acknowledgments.} The authors would like to thank 
Nuria Santervás Arranz, Sanaz Gerivani, Noor Ul Ain and Hongrui Zhang 
for useful discussions.
  This project is partially supported by the European Commission Horizon 
MSCA-SE Project MAMBA (Grant No. 101131245).
  RZ acknowledges funding from the National Natural Science 
Foundation of China (52506192) and the Fundamental Research Funds 
for the Central Universities (xzy012025011).
  EA acknowledges funding from the Spanish MCIN/AEI/10.13039/501100011033
(through grant PID2022-139776NB-C65, and a Mar\'{\i}a de Maeztu award to 
Nanogune, Grant CEX2020-001038-M), the United Kingdom's EPSRC 
(Grant EP/V062654/1), and the Basque Ikur HPC program, through a 
2023 Cotutelage grant.
  We also acknowledge computational time provided by the Spanish
HPC Network, RES, in the Mare Nostrum V computer at the Barcelona
Supercomputing Centre, through grants FI-2025-3-0062 
\& FI-2026-1-0071.


%

\appendix

\renewcommand{\thefigure}{A\arabic{figure}} 
\setcounter{figure}{0}

\section*{\label{app.method}Appendix}

\begin{figure}[htbp]
    \centering
    \includegraphics[width=1\linewidth]{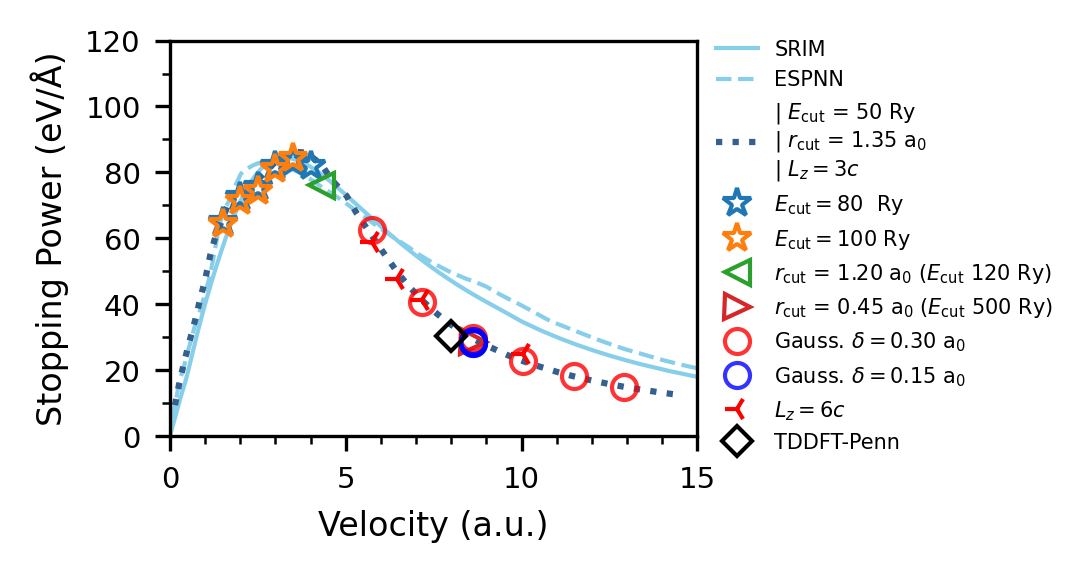}
    \caption{\label{fig-a1} Verification of numerical robustness of the high-velocity 
    (\(v \gtrsim 5\) a.u.) stopping power in off-channeling. 
    Calculations using different plane-wave cutoffs, pseudopotential cutoff radii, 
    Gaussian point-charge widths, and an enlarged simulation cell show no 
    significant variation.} 
\end{figure}

  \textit{Method Details.} In the high-velocity regime (\( v \gtrsim 5\) a.u.) 
and beyond the Bragg peak, the calculated \(\mathcal{S}_\text{e}\) is lower 
than the prediction of the empirical models, even along the off-channeling 
trajectory. 
  This deviation is expected to vanish for \(v \gtrsim 20 \) a.u., based on the 
observed trend in Fig.~\ref{fig-a1}.


  We systematically examined the influence of numerical and methodological 
factors. 
  Specifically, we tested a range of higher plane-wave kinetic energy cutoffs 
\(E_\text{cut}\) and smaller pseudopotential cutoff radii \(r_\text{cut}\). 
  We also replaced pseudopotentials by local potentials generated
by spherical Gaussian distributions representing the 
nuclear point charges for the projectile and host, using different 
Gaussian-width parameters \(\delta\).
  We thereby eliminated potential effects associated with non-local 
pseudopotentials~\cite{stengel2026rototranslational} and with the regularization 
of the Coulomb singularity at the nuclear position. 
  We also employed a larger simulation cell with \(L_z = 6c\) to assess the 
influence of system size.
  As shown in Fig.~\ref{fig-a1}, none of these tests leads to a significant 
change in the calculated stopping power.
  Furthermore, calculations of Be ion in two jellium hosts, dilute 
(valence-electron-equivalent density) and dense (all-electron-equivalent density), 
provide theoretical bounds (see Fig.~\ref{fig1} in the main text). 
  The dilute jellium results are consistent with hyper-channeling, with off-channeling 
converging to it at higher velocities. 
  Interestingly, the dense jellium results align with the empirical models. 
Since the full delocalization of all electrons in Be (including the 1s)
is hardly an accurate description of the system, the latter agreement offers 
a plausible explanation of the discrepancy of our Be in Be results with the models. 
  To further support this, the TDDFT-Penn 
approach~\cite{matias2025stopping} is employed, which models the real material 
as a combination of jelliums with various resonant frequencies and 
assigns weights based on the energy loss function~\cite{chen2022influence}, which is expected to be accurate
at high velocities. 
  The results obtained from this additional TDDFT-Penn calculation 
(Fig.~\ref{fig-a1}) are consistent with our calculations in 
the high-velocity regime.
  
  The above extension tests demonstrate the consistency of the calculated results. 
  The systematic deviation in the high-velocity regime may arise from two 
remaining factors. 
  On the one hand, TDDFT may have limitations in treating multi-shell systems 
at high velocities.
  For example, there may be issues with the currently used TDDFT functionals 
(based on the adiabatic approximation and lacking memory effects), although the 
fact that the same functional agrees wit the models when the host density is 
spread as jellium seems to point away from the exchange-correlation functional.
  On the other hand, empirical models such as SRIM also have their own 
limitations, which exhibit deviations when predicting stopping power in the 
presence of heavy elements~\cite{ziegler2010srim,ullah2018core}. 
  However, this discrepancy does not affect our discussion of the role 
of core electrons in the main text and we leave it as an open question for 
future work.

\begin{figure}[htbp]
    \centering
    \includegraphics[width=0.95\linewidth]{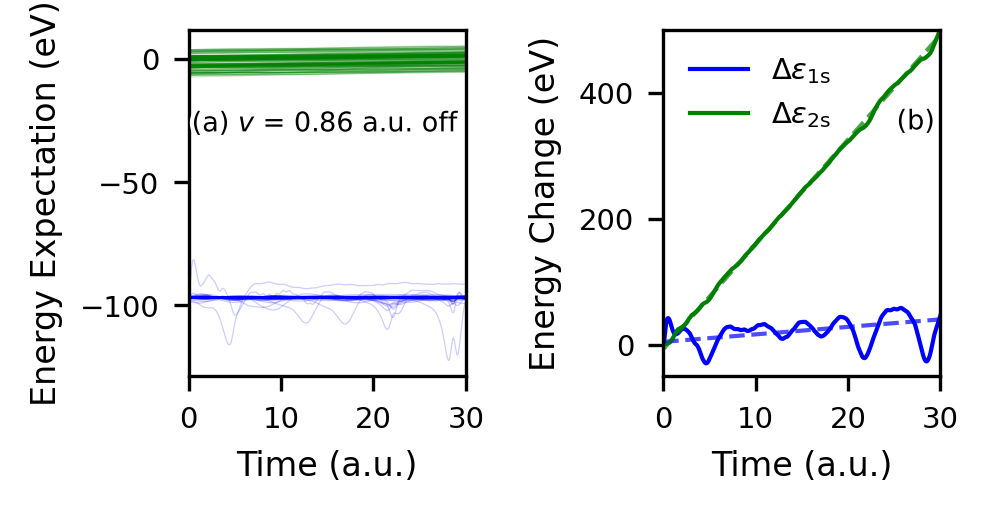}

    \caption{\label{fig-a2} (a) Instantaneous energy expectation values of KS 
    wavefunctions along the off-channeling trajectory at \(v = 0.86\) a.u. 
    Green lines correspond to initially 2$s$ states, and blue lines to initially 
    1$s$ states. 
    (b) Solid lines show the instantaneous sum of the energy expectations relative 
    to their initial values, while the dashed lines indicate linear fits used to 
    extract the corresponding stopping component.
} 
\end{figure}

    \textit{Energy.} We analyzed the individual energy expectation 
values of all KS wavefunctions system as a function of the moving projectile’s 
propagation time, \(\varepsilon_i =  \bra{\psi_i(t)} \hat{h}_\text{KS}(t) 
\ket{\psi_i(t)}\) (see Fig.~\ref{fig-a2}(a)). 
  The time-dependent sum of the energy expectation values of the 1$s$ 
band (blue solid line) and the 2$s$ band (green solid line) are shown 
in Fig.~\ref{fig-a2}(b).

  \textit{Energy Excitation Distribution.} To directly visualize the 
electronic excitations responsible for energy loss, we analyze the 
time-evolving electronic excitation distribution~\cite{zeb2012electronic}. 
  By projecting the evolving states $|\psi_n(t)\rangle$ onto the basis 
of adiabatic states \(|\phi_i\rangle \), \( C_{n,i}(t) = \langle \phi_i 
| \psi_n(t) \rangle \), we construct the time-dependent occupied energy 
distribution \(O(E, t) = \sum_{n} f_n \sum_{i} |C_{n,i}(t)|^2 \, 
\delta(E - \epsilon_i),\) where \( \epsilon_i \) is the KS eigenvalue 
of the adiabatic state \( |\phi_i\rangle \). 
  The electronic excitation distribution $P(E)$ due to projectile passage 
is then obtained by subtracting from $O(E)$ the 
ground-state electronic density of states, \( g(E) \),
\(P(E, t) = O(E, t) - \Theta(E_F - E) \, g(E)\), where \( E_F \) is the Fermi 
energy and \( \Theta \) is the Heaviside step function. 
  \( P(E,t) < 0 \) indicates depopulation (holes) below \( E_F \), and 
\( P(E,t) > 0 \) indicates population (electrons) above \( E_F \). 
  The numbers of excited electrons are estimated as \(\mathcal{N}_{1s}(t) 
= \frac{1}{vt} \int_{-\infty}^{E_{\text{gap}}} |P(E)| \, dE\) and 
\(\mathcal{N}_{2s}(t) = \frac{1}{vt} \int_{E_{\text{gap}}}^{E_F} |P(E)| \, dE\), 
where \(E_\text{gap} = -90\) eV is a reference energy chosen within the 
gap separating the 1$s$ and the 2$s$ bands.

\end{document}